\documentclass{article} 
\usepackage[final]{colm2026_conference}

\usepackage{microtype}
\usepackage{hyperref}
\usepackage{url}
\usepackage{booktabs}
\usepackage{multirow}
\usepackage{amsmath}
\usepackage{xcolor}

\usepackage{listings}
\usepackage{floatrow}
\usepackage{placeins}
\usepackage{capt-of}

\usepackage{lineno}

\usepackage{graphicx}
\usepackage[T1]{fontenc}

\usepackage[most]{tcolorbox}
\usepackage{xcolor}
\usepackage{varwidth}
\usepackage{array}
\usepackage{ragged2e}
\usepackage{wrapfig}
\usepackage{makecell}
\usepackage{pifont}

\usepackage{titletoc}

\definecolor{MCQDark}{HTML}{3F78A8}
\definecolor{MCQMid}{HTML}{6C97BE}
\definecolor{MCQLight}{HTML}{EAF2FA}

\definecolor{CodeDark}{HTML}{C6962B}
\definecolor{CodeMid}{HTML}{D7AC4B}
\definecolor{CodeLight}{HTML}{F8EFCF}

\definecolor{PanelBorder}{HTML}{A8B7CC}
\definecolor{PanelBg}{HTML}{F7F8FA}
\definecolor{SoftText}{HTML}{4B5A70}

\definecolor{ResolveBlue}{HTML}{3F78A8}
\definecolor{ScaffoldGold}{HTML}{D7AC4B}
\definecolor{ContentBrown}{HTML}{9B6742}
\definecolor{PanelBorder}{HTML}{A8B7CC}
\definecolor{PanelBg}{HTML}{F7F8FA}

\definecolor{darkblue}{rgb}{0, 0, 0.5}
\definecolor{codebg}{RGB}{247, 247, 247}
\definecolor{codekeyword}{RGB}{0, 66, 154}
\definecolor{codecomment}{RGB}{0, 128, 0}
\definecolor{codestring}{RGB}{163, 21, 21}
\hypersetup{colorlinks=true, citecolor=darkblue, linkcolor=darkblue, urlcolor=darkblue}

\newcommand{\cm}{\text{\ding{51}}}
\newcommand{\xm}{\text{\ding{55}}}

\lstdefinestyle{paperpython}{
  language=Python,
  basicstyle=\ttfamily\footnotesize,
  keywordstyle=\color{codekeyword}\bfseries,
  commentstyle=\color{codecomment},
  stringstyle=\color{codestring},
  showstringspaces=false,
  breaklines=true,
  columns=fullflexible,
  keepspaces=true,
  frame=single,
  framerule=0.3pt,
  rulecolor=\color{black!20},
  backgroundcolor=\color{codebg},
  xleftmargin=0.8em,
  framexleftmargin=0.6em,
  framesep=4pt,
  aboveskip=0.6em,
  belowskip=0.6em
}

\newtcolorbox{NewBoxFloat}[3][ResolveBlue]{%
  enhanced,
  width=\linewidth,
  colback=PanelBg,
  colframe=PanelBorder,
  colbacktitle=#1,
  coltitle=white,
  fonttitle=\bfseries,
  boxrule=0.9pt,
  arc=1.5mm,
  left=0.6em, right=0.6em,
  top=0.5em, bottom=0.5em,
  leftupper=0.4em, rightupper=0.4em,
  title={#2},
  label={#3},
}

\usepackage{lineno}

\definecolor{darkblue}{rgb}{0, 0, 0.5}
\hypersetup{colorlinks=true, citecolor=darkblue, linkcolor=darkblue, urlcolor=darkblue}

\title{Revision or Re-Solving? \\ Decomposing Second-Pass Gains in Multi-LLM Pipelines}

\author{
Jingjie Ning\thanks{These authors contributed equally to this work.} \qquad
Xueqi Li\footnotemark[1] \qquad
Chengyu Yu\footnotemark[1] \\
\vspace{0.1em} \\
School of Computer Science, Carnegie Mellon University \\
\texttt{\{jening, xueqil, cy4\}@andrew.cmu.edu}
}

\begin{document}
 
\ifcolmsubmission
\linenumbers
\fi

\maketitle

\begin{abstract}
Multi-LLM revision pipelines use one model to review another model's draft and are often described as collaborative error correction. We study what produces their second-pass gain. A controlled four-condition design separates the gain into re-solving, scaffold, and content effects. We evaluate two heterogeneous model pairs on three benchmarks spanning knowledge-intensive MCQ and competitive programming. On MCQ tasks, most gains are consistent with stronger-model re-solving, and direct routing to the stronger model can outperform revision of a weak draft. On code generation, two-stage prompting remains useful because semantically null drafts provide substantial structural scaffolding, while weak draft content can be harmful. Role-reversed experiments show that strong drafts benefit weak reviewers. The same task-dependent patterns recur in both model pairs. The controls also apply when one model or agent passes an intermediate artifact to another. They distinguish information transfer from downstream recomputation and interface effects, which makes them relevant to routing, communication, resource allocation, and graph design in model cascades and multi-agent workflows.

\end{abstract}

\section{Introduction}
\label{sec:intro}

\begin{wrapfigure}{r}{0.52\columnwidth}
\vspace{-8pt}
    \centering
    \includegraphics[width=1.0\columnwidth]{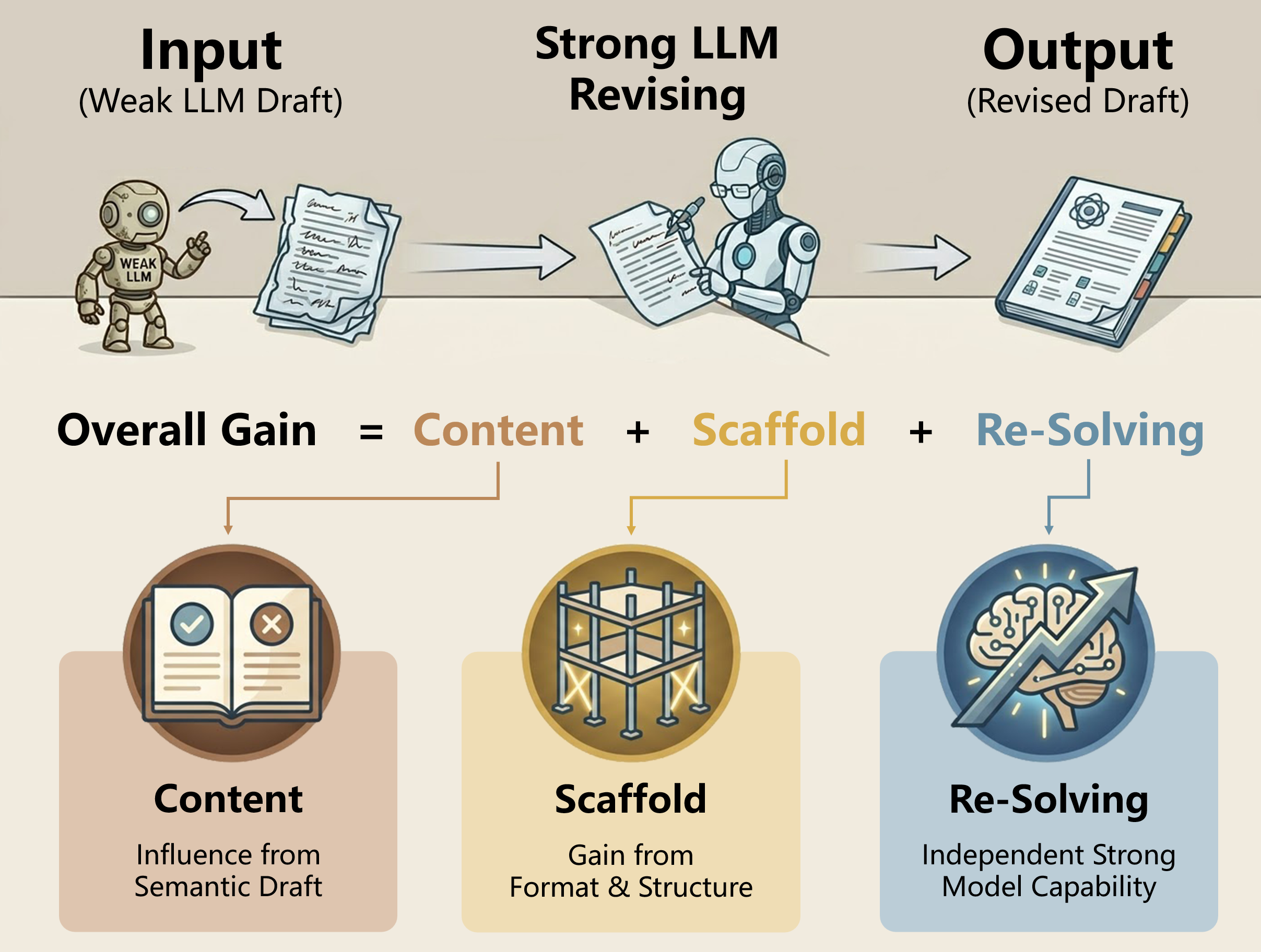}
    \caption{Illustration of the decomposition.}
    \label{fig:decomposition_overview}
    \vspace{-8pt}
\end{wrapfigure}

\vspace{-4pt}

Using one language model to critique and refine another model's output has become a common way to improve response quality beyond single-pass generation~\citep{madaan2023selfrefine,shinn2023reflexion,chen2024selfdebug}. The same handoff pattern increasingly appears in model cascades and multi-agent workflows, where planners, solvers, critics, and verifiers exchange intermediate artifacts. The usual intuition is that downstream models improve the artifact by exploiting information contributed upstream, and end-to-end results often appear consistent with this view.

End-to-end improvement alone, however, cannot reveal whether the handoff transferred useful information. When the reviewer is also the stronger model, an apparent revision gain may arise simply because it can independently re-solve the problem. The consequential question is therefore not whether an additional model call helps, but 
\emph{what resource supplied by that call produces the gain}. If gains mostly come from re-solving, routing the query to the stronger model may be enough; if they come from scaffold or content, multi-stage collaboration remains useful. This distinction turns aggregate accuracy into an architectural diagnosis and may differ sharply across task types.

Prior work has largely measured overall revision gains without separating these mechanisms~\citep{madaan2023selfrefine,paul2024refiner,olausson2023self}. The few studies that include re-solving controls are narrower in scope and do not test role-reversed settings or separate the structural and semantic contributions of the draft in a unified controlled design~\citep{stechly2023gpt4,huang2024large}.

We introduce a four-condition design that additively decomposes second-pass gains into three effects: re-solving, scaffold, and content, as shown in Figure~\ref{fig:decomposition_overview}. We evaluate this design on two model pairs across GPQA Diamond~\citep{rein2023gpqa}, HLE~\citep{phan2025hle}, and LiveCodeBench~\citep{jain2025livecodebench}, with an additional role-reversed setting. Because the design requires only an upstream artifact and a downstream model, it also provides a reusable unit of analysis for substantially broader forms of sequential collaborative inference.

Our main findings are:
\begin{enumerate}
  \item On knowledge-intensive MCQ tasks, the content effect is non-significant across both model pairs and datasets, and most observed gains are consistent with stronger-model re-solving.
  \item On code generation, weak-draft content is significantly worse than a structure-only scaffold. Gains are driven mainly by scaffold, and the content disadvantage grows with problem difficulty.
  \item In the role-reversed setting, strong-model drafts significantly help weaker reviewers on both MCQ and code tasks, suggesting that draft utility depends on draft quality and on the kind of information the draft provides.
  \item Across both heterogeneous model pairs, these task-dependent regimes recur, motivating a general diagnostic framework for model cascades and multi-agent systems that distinguishes transferred semantics from structural coordination and additional downstream computation.
\end{enumerate}

\section{Related work}
\label{sec:related}

Revision-style inference and self-correction are widely used to improve large language model (LLM) outputs beyond single-shot generation. Our work sits at the intersection of (i) iterative refinement and critique pipelines, (ii) analyses questioning when self-correction truly works, and (iii) multi-agent and capability-asymmetric collaboration.

\vspace{-5pt}
\paragraph{Iterative refinement with language feedback.}
Self-Refine~\citep{madaan2023selfrefine} popularizes a feedback$\rightarrow$refine loop in which an LLM critiques and iteratively revises its own output without additional training. Reflexion~\citep{shinn2023reflexion} extends this to agentic settings by storing verbal reflections in memory for future attempts. SCREWS~\citep{shridhar2023screws} organizes revision around modular sampling, conditional-resampling, and selection strategies, while REFINER~\citep{paul2024refiner} trains a critic to provide structured feedback over intermediate reasoning representations. DeCRIM~\citep{ferraz2024decrim} applies a decompose--critique--refine pattern to multi-constraint instruction following. In programming, Self-Debugging~\citep{chen2024selfdebug} uses execution results to enable iterative code repair, and CRITIC~\citep{gou2023critic} interleaves generation with tool-based validation. These approaches design increasingly capable revision procedures; our focus is the complementary attribution question of which part of a revision procedure actually produces its gain.

\vspace{-5pt}
\paragraph{Limits of intrinsic self-correction.}
A growing line of work argues that intrinsic self-correction, revision without reliable external feedback, is often ineffective and may even degrade performance~\citep{huang2024large}. \citet{stechly2023gpt4} show that apparent gains from iterative prompting can be explained by sampling and selection effects rather than by critique content. Surveys conclude that self-correction works best with credible external feedback or explicit correction training~\citep{kamoi2024survey,pan2024autocorrect}. Related analyses show that confidence shapes revision outcomes and that, in continual agent skill learning, external feedback can yield iterative improvement whereas self-feedback can induce recursive drift~\citep{li2024confidence,zhong2026skilllearnbench}. Together, these results support treating ``revision'' as a composition of distinct effects rather than a single mechanism.

\vspace{-5pt}
\paragraph{Draft quality, capability asymmetry, and multi-agent decomposition.}
In code generation, self-repair gains are often modest, underscoring that editing low-quality drafts is not uniformly beneficial~\citep{olausson2023self}. Multi-agent debate frameworks improve reasoning and factuality~\citep{du2023debate,liang2024mad}, but recent decompositions reveal simple ensembling accounts for much of the observed gain~\citep{choi2025debateorvote}. Weak-to-strong generalization studies find stronger models benefit from weak supervision but may still underperform strong ceilings~\citep{burns2023weaktostrong}. These works expose individual explanations for collaborative gains; our framework places re-solving, scaffold, and content in one matched accounting design, enabling their relative contributions to be compared across tasks, model pairs, and role directions.

\vspace{-5pt}
\paragraph{Re-solving and test-time scaling.}
Recent test-time scaling work shows that abandoning a flawed trajectory and re-solving can outperform patching it~\citep{wang2026re2,shi2025ssr}. Parallel agentic-search rollouts can likewise become redundant when shared initial queries anchor their retrieved evidence; diverse initialization improves exploration at matched compute~\citep{murali2026divinit}. Our MCQ result aligns with this literature, while scaffold and content isolate what remains beyond fresh or parallel solving.

\section{Method}
\label{sec:method}

As discussed in Section~\ref{sec:intro}, the total second-pass gain $x_2 - x_1$ confounds model capability, review framing, and draft semantics. We introduce a four-condition design that decomposes this gain into three interpretable components.

\subsection{Four Conditions}

We define four conditions, $x_1$--$x_4$, and use these symbols to refer to both the conditions and their corresponding accuracies. We evaluate two model pairs, \textbf{Pair~1} (Gemini Flash Lite~$\to$~GPT-5-mini) and \textbf{Pair~2} (GPT-4o-mini~$\to$~Gemini Flash), each in a \textbf{Primary} setting (weak~$\to$~strong) and a \textbf{Supplementary} setting (roles swapped).

\textbf{$x_1$: Generator baseline.} The generator answers the question directly. The result is cached and reused by all downstream conditions.

\textbf{$x_2$: Standard cross-model revision.} The reviewer receives the question together with the generator's output and is asked to review and improve the answer. The prompt template is identical to~$x_4$; only the draft argument differs.

\textbf{$x_3$: Re-solving control.} The reviewer receives only the question and answers it directly, using the same prompt as~$x_1$ with no review framing. Any framing benefit from the review task therefore accrues to~$x_2$, making $x_2 > x_3$ a conservative test of draft utility.

\textbf{$x_4$: Scaffold control.} The reviewer receives the question together with a semantically null placeholder draft (described in Section~\ref{sec:nulldraft}), using the same review prompt as~$x_2$.

No prompt in any condition reveals the identity of the model that produced the draft. For MCQ, all conditions use the format ``Reason~1 / Reason~2 / Answer:~X'', placing the answer letter last to prevent it from anchoring the stated reasons.

\subsection{Effect Decomposition}

Using the four conditions, the total second-pass gain $x_2 - x_1$ decomposes additively as:

\begin{figure}[H]
\vspace{-5pt}
    \centering
    \includegraphics[width=0.7\linewidth]{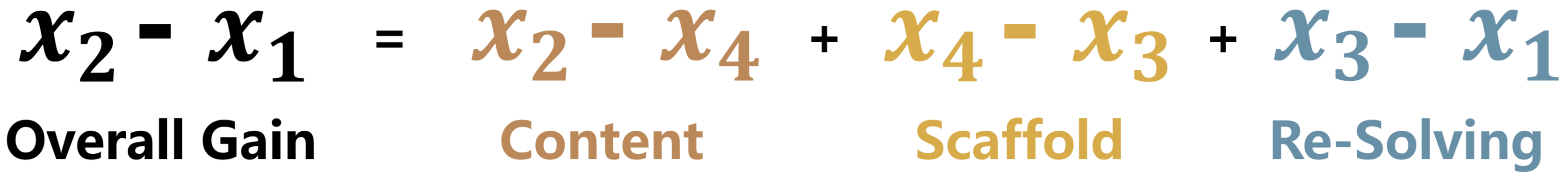}
    \label{fig:main_formula}
    \vspace{-15pt}
\end{figure}
The \textbf{content effect}, which is our primary quantity of interest, measures the marginal value of the generator's actual output relative to a format-matched null draft. A negative content effect indicates that the incoming draft misleads the reviewer. The \textbf{scaffold effect} captures the incremental value of the review framing and structural template, while the \textbf{re-solving effect} measures the reviewer--generator capability difference under direct-answer framing.

Importantly, this accounting does not require re-solving, scaffold, and content to be statistically independent. Instead, it gives an exact operational account of three counterfactual transitions within a fixed revision pipeline. Their magnitudes are allowed to vary jointly with task structure, reviewer capability, and draft quality; such interactions appear as changes in the marginal contributions across tasks, model pairs, and role directions rather than as an unassigned residual. In particular, the role-reversed setting probes how the full decomposition changes when draft quality and reviewer capability trade places. The framework therefore characterizes both the contribution of each pipeline component and the conditions under which that contribution changes sign or scale.

\subsection{Null Draft Design}
\label{sec:nulldraft}

The null draft in $x_4$ preserves the format of a genuine $x_1$ response while removing task-specific content. We do not use another question's real output as a decoy, because it may carry latent transfer signal. A semantically null but well-formed draft avoids such transfer while still eliciting normal reviewer behavior. For MCQ tasks, we use a generic two-reason template followed by \texttt{Answer: X}, where \texttt{X} \(\in\) \{A, B, C, D\}. For code tasks, the null draft is a syntactically valid but semantically empty stub. The full null draft is shown in Appendix~\ref{app:exp_setup}.

\subsection{Datasets and Evaluation}

We evaluate on three benchmarks. \textbf{GPQA Diamond}~\citep{rein2023gpqa} (198 questions) consists of graduate-level multiple-choice science questions in physics, chemistry, and biology. \textbf{HLE} (Humanity's Last Exam)~\citep{phan2025hle} (451 questions) contains expert-level questions across many disciplines and serves as a higher-difficulty MCQ complement to GPQA. \textbf{LiveCodeBench}~\citep{jain2025livecodebench} (1,054 problems) provides competitive-programming problems released after model training cutoffs, reducing contamination concerns. Problems are partitioned into Easy (322), Medium (382), and Hard (350) using the benchmark's own difficulty labels, and solutions are evaluated against all public test cases. All statistical tests use two-tailed McNemar's test with Yates continuity correction at $\alpha = 0.05$.

\vspace{-5pt}

\section{Results and Analysis}
\label{sec:results}

\begin{figure}[t]
\vspace{-10pt}
    \begin{center}
    \includegraphics[width=1.0\linewidth]{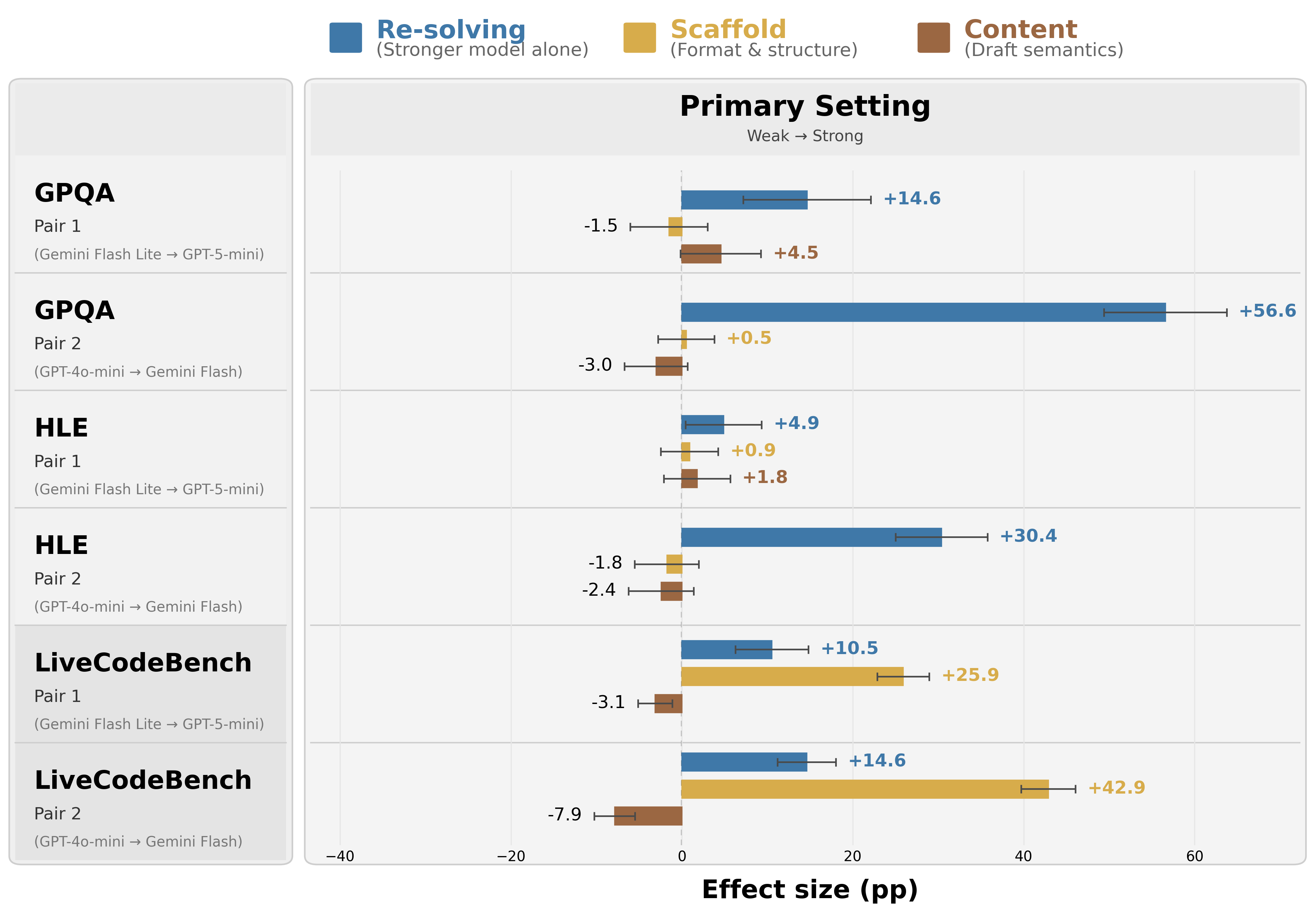}
    \end{center}
    \caption{Signed decomposition of second-pass gains in the (weak $\rightarrow$ strong) setting into re-solving, scaffold, and content. MCQ gains are mostly re-solving, whereas LiveCodeBench is scaffold-dominated with negative content. Error bars show paired 95\% CI.}
    \label{fig:primary_decomposition}
    \vspace{-10pt}
\end{figure}

Figure~\ref{fig:primary_decomposition} gives the global picture for the primary setting, our main setting of interest. Second-pass gains are not monolithic. On MCQ, they are dominated by re-solving, whereas on LiveCodeBench they are dominated by scaffold, with weak-draft content becoming harmful. We discuss the two task types separately and then use the supplementary strong$\rightarrow$weak setting to examine how the complete decomposition transforms when the direction of the capability asymmetry is reversed.

\vspace{-5pt}

\subsection{MCQ tasks}

Across both datasets, MCQ behavior is best understood as re-solving-dominated in the primary direction and draft-quality-dependent in the supplementary direction.

\subsubsection{Primary Setting}

Table~\ref{tab:mcq_combined} shows a consistent pattern across GPQA and HLE. In the weak$\rightarrow$strong direction, most of the second-pass gain is already captured by re-solving, while scaffold remains negligible and content stays near zero. The clearest examples are GPQA Pair~2 and HLE Pair~2, where re-solving contributes $+56.6$ pp and $+30.4$ pp, respectively, while content remains small and non-significant. Even when revision substantially improves over the weak baseline, the gain is therefore largely consistent with the stronger reviewer solving the question independently rather than extracting reliable semantic value from the weak draft.

Mechanistically, this pattern is plausible for knowledge-intensive MCQ tasks because the answer space is constrained and the draft provides little task-specific structural guidance. A stronger reviewer can often discard a weak rationale and reconstruct the answer from its own parametric knowledge, so the draft functions less as a useful intermediate object than as a noisy prefix. This also explains why content remains near zero rather than consistently negative. The weak draft is not always inert, but its helpful and harmful influences mostly wash out, leaving re-solving as the dominant source of gain.

\begin{table}[t]
    \footnotesize
    \centering
    \renewcommand{\arraystretch}{1.1}
    \caption{Accuracy and effect decomposition on MCQ tasks. The \textbf{primary} setting (weak $\to$ strong) reports accuracy and all three effects; the \textbf{supplementary} setting (strong $\to$ weak) reports effects only (full accuracy in Appendix Table~\ref{tab:full_accuracy}). $^{***}\!p<.001$; $^{*}\!p<.05$; ns: not significant (two-tailed McNemar's test with Yates correction).}
    \label{tab:mcq_combined}
    \setlength{\tabcolsep}{2.4pt}
    \begin{tabular}{@{}llccccccc!{\hspace{6pt}\vrule width 0.4pt\hspace{6pt}}ccc@{}}
    \toprule
    & & \multicolumn{7}{c!{\hspace{6pt}\vrule width 0.4pt\hspace{6pt}}}{\textbf{Primary}} 
    & \multicolumn{3}{c}{\textbf{Supplementary}} \\
    \cmidrule(lr){3-9}\cmidrule(l){10-12}
    & & \multicolumn{4}{c}{Accuracy (\%)} & \multicolumn{3}{c!{\hspace{6pt}\vrule width 0.4pt\hspace{6pt}}}{Effects (pp)} 
    & \multicolumn{3}{c}{Effects (pp)} \\
    \cmidrule(lr){3-6}\cmidrule(lr){7-9}\cmidrule(l){10-12}
    Dataset & Pair & $x_1$ & $x_2$ & $x_3$ & $x_4$ & Re-solv. & Scaff. & Content & Re-solv. & Scaff. & Content \\
    \midrule
    \multirow{2}{*}{GPQA}
      & P1 & 61.6 & 79.3 & 76.3 & 74.7 
           & +14.6$^{***}$ & $-$1.5$^{\mathrm{ns}}$ & +4.5$^{\mathrm{ns}}$  
           & $-$14.6$^{***}$ & $-$5.1$^{\mathrm{ns}}$ & +14.6$^{***}$ \\
      & P2 & 33.3 & 87.4 & 89.9 & 90.4 
           & +56.6$^{***}$ & +0.5$^{\mathrm{ns}}$  & $-$3.0$^{\mathrm{ns}}$ 
           & $-$55.6$^{***}$ & $-$3.5$^{\mathrm{ns}}$ & +26.8$^{***}$ \\
    \midrule
    \multirow{2}{*}{HLE}
      & P1 & 12.6 & 20.2 & 17.5 & 18.4 
           & +4.9$^{*}$     & +0.9$^{\mathrm{ns}}$  & +1.8$^{\mathrm{ns}}$  
           & $-$3.8$^{\mathrm{ns}}$  & $-$0.7$^{\mathrm{ns}}$ & +2.9$^{\mathrm{ns}}$ \\
      & P2 & 11.3 & 37.5 & 41.7 & 39.9 
           & +30.4$^{***}$  & $-$1.8$^{\mathrm{ns}}$ & $-$2.4$^{\mathrm{ns}}$ 
           & $-$30.6$^{***}$ & $-$1.1$^{\mathrm{ns}}$ & +12.0$^{***}$ \\
    \bottomrule
    \end{tabular}
\end{table}

\subsubsection{Supplementary Setting}

The supplementary strong$\rightarrow$weak results reverse this picture. Re-solving becomes negative because the independent second pass is now performed by the weaker model, scaffold remains small, and content becomes the principal compensating contribution. Content is significantly positive in both GPQA pairs and in HLE Pair~2, while HLE Pair~1 remains weak and non-significant. In GPQA and HLE Pair~2, the stronger generator produces drafts that are good enough for the weaker reviewer to reuse; in HLE Pair~1, the supplementary generator reaches only 18.0\% accuracy, leaving little reliable signal to transfer. The full decomposition therefore supports a threshold-like view of draft utility. Weak drafts add little beyond what a stronger reviewer can do on its own, whereas sufficiently strong drafts can transfer capability to a weaker reviewer.

\begin{figure}[h]
    \centering
    \includegraphics[width=0.95\linewidth]{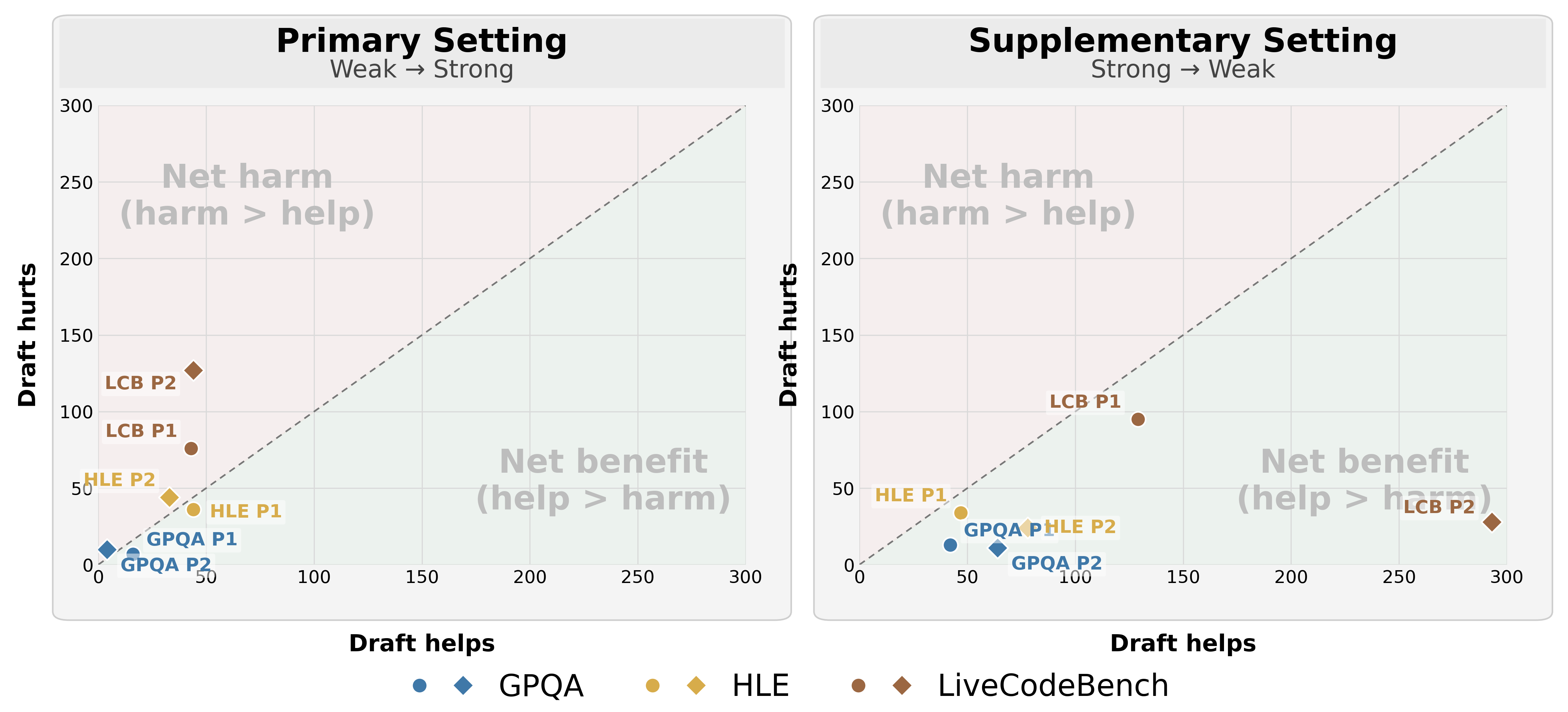}
\caption{Benefit--harm view of the real-draft content effect ($x_2-x_4$). Points below the diagonal indicate net benefit and points above it net harm. MCQ lies near cancellation, whereas LiveCodeBench shifts from net harm to net benefit across settings.}
    \label{fig:benefit_harm_cancellation}
\end{figure}

\subsubsection{Benefit--Harm View of the Content Effect}
Figure~\ref{fig:benefit_harm_cancellation} provides a complementary view of the content effect by separating draft benefits from draft harms. On MCQ, the primary-setting points lie near the diagonal, consistent with near-cancellation. On LiveCodeBench, they shift clearly toward net harm, while the supplementary setting moves toward net benefit. We next show that this asymmetry is sharper on code generation.

\subsection{Code Generation}

Code generation exhibits a different regime from MCQ. In the primary direction, the dominant gain comes from scaffold rather than from re-solving alone, and weak-draft content becomes actively harmful.

\subsubsection{Primary Setting}

\begin{table}[t]
    \footnotesize
    \centering
    \renewcommand{\arraystretch}{1.1}
    \caption{Accuracy and effect decomposition on LiveCodeBench. 
    The \textbf{primary} setting (weak $\to$ strong) reports accuracy and all three decomposed effects; 
    the \textbf{supplementary} setting (strong $\to$ weak) reports effects only 
    (full accuracy in Appendix Table~\ref{tab:full_accuracy}).
    $^{***}\!p<.001$; $^{**}\!p<.01$; $^{*}\!p<.05$; ns: not significant.
    }
    \label{tab:code_combined}
    \setlength{\tabcolsep}{2.4pt}
    \begin{tabular}{@{}lcccccccc!{\hspace{6pt}\vrule width 0.4pt\hspace{6pt}}ccc@{}}
    \toprule
    & \multicolumn{8}{c!{\hspace{6pt}\vrule width 0.4pt\hspace{6pt}}}{\textbf{Primary}} 
    & \multicolumn{3}{c}{\textbf{Supplementary}} \\
    \cmidrule(lr){2-9}\cmidrule(l){10-12}
    & \multicolumn{4}{c}{Accuracy (\%)} & \multicolumn{4}{c!{\hspace{6pt}\vrule width 0.4pt\hspace{6pt}}}{Effects (pp)} 
    & \multicolumn{3}{c}{Effects (pp)} \\
    \cmidrule(lr){2-5}\cmidrule(lr){6-9}\cmidrule(l){10-12}
    Pair & $x_1$ & $x_2$ & $x_3$ & $x_4$ & Total & Re-solv. & Scaffold & Content & Re-solv. & Scaffold & Content \\
    \midrule
    Pair 1 
      & 50.6 & 83.9 & 61.1 & 87.0 
      & +33.3 & +10.5$^{***}$ & +25.9$^{***}$ & $-$3.1$^{**}$ 
      & $-$7.3$^{**}$ & +9.2$^{***}$ & +3.2$^{*}$ \\
    Pair 2 
      & 28.5 & 78.1 & 43.1 & 86.0 
      & +49.6 & +14.6$^{***}$ & +42.9$^{***}$ & $-$7.9$^{***}$ 
      & $-$15.6$^{***}$ & $-$1.7$^{\mathrm{ns}}$ & +25.1$^{***}$ \\
    \bottomrule
    \end{tabular}
\end{table}
    
Table~\ref{tab:code_combined} shows that the core pattern on LiveCodeBench is not re-solving-dominated but scaffold-dominated. In both model pairs, the null scaffold outperforms the standard revision pipeline, and the content effect is significantly negative. This means that once the reviewer is given a weak code draft, the problem is no longer just whether the stronger model can solve the task, but whether it can avoid being pulled into a bad intermediate artifact.

\noindent
\begin{minipage}[t]{0.45\linewidth}
\vspace{0pt}
A plausible mechanism is artifact-level anchoring. Unlike MCQ, where the draft mainly provides a reasoning prefix, code drafts expose a partially instantiated executable object including function signatures, parsing logic, wrapper structure, and implementation choices. Even when the weak model's high-level algorithm is not entirely wrong, these concrete implementation decisions can still trap the reviewer in a brittle local trajectory. The strong reviewer therefore behaves less like an independent solver and more like an editor working within the weak draft's structural constraints, so weak-draft content can actively hurt performance even when the second pass remains beneficial.

\smallskip
Figure~\ref{fig:lcb_difficulty} reinforces this interpretation. For Pair~1, the content effect becomes increasingly negative from easy to medium and hard. As difficulty rises, weak drafts are more likely to contain deeply flawed implementation structure, increasing the anchoring cost.
\end{minipage}\hfill
\begin{minipage}[t]{0.51\linewidth}
\vspace{0pt}
    \centering
    \includegraphics[width=0.96\linewidth]{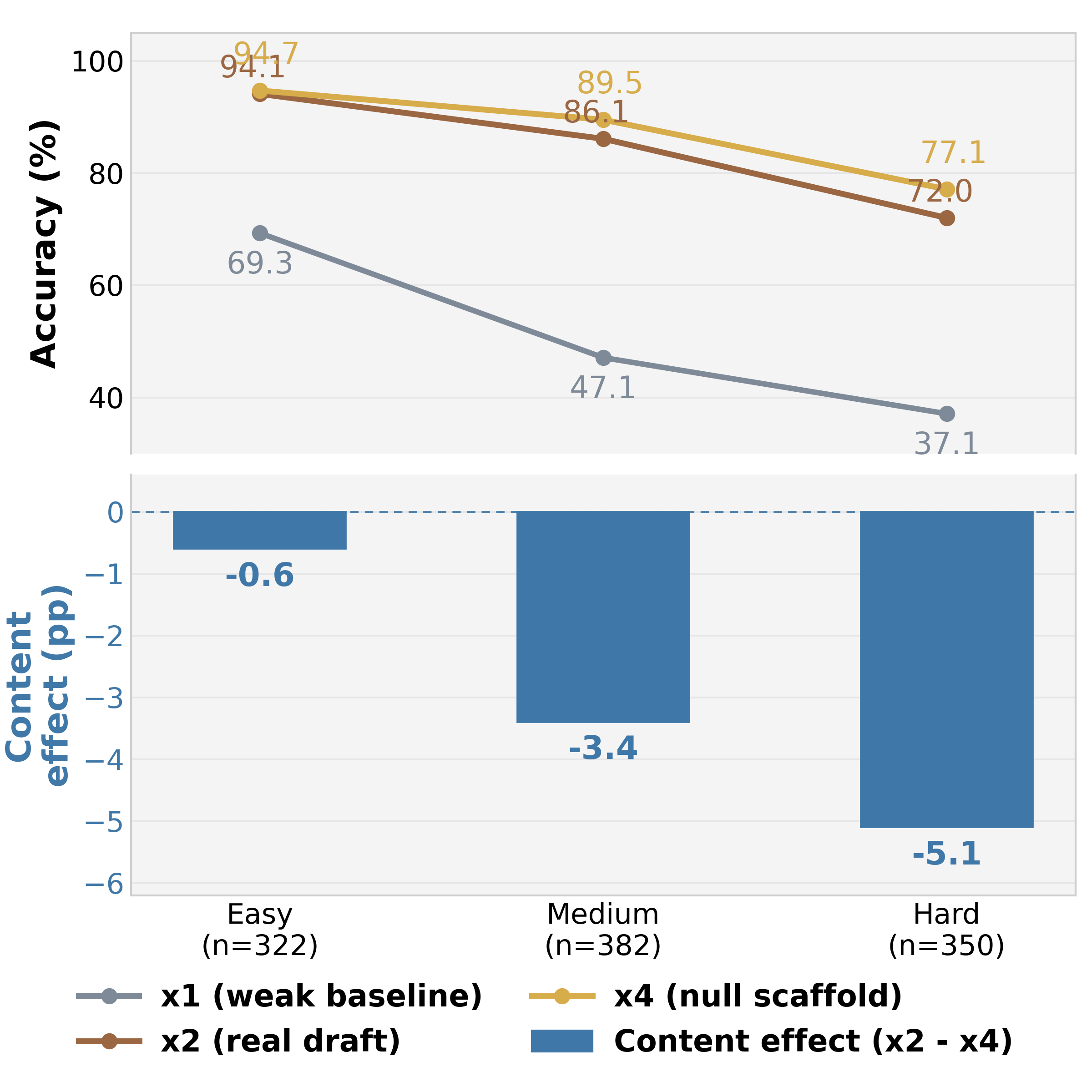}
    \captionof{figure}{LiveCodeBench difficulty split for Pair~1. The content effect becomes increasingly negative from easy to hard.}
    \label{fig:lcb_difficulty}
\end{minipage}

\vspace{3pt}

The scaffold effect is therefore not just a statistical residual but part of the mechanism. A null code draft provides no algorithmic solution, yet it still offers a syntactically valid, code-shaped object that waits for completion and repair. In the review framing, this appears to shift the second pass away from free-form solving and toward submission repair that fills in wrappers, normalizes I/O structure, and produces a complete executable artifact. Unlike MCQ, code tasks can benefit from this kind of structural intermediate even when its semantic content is empty. An appendix ablation comparing the \textbf{Null Scaffold} and the \textbf{True-Null Scaffold} shows only negligible differences (Appendix Table~\ref{tab:scaffold_ablation}), suggesting that these scaffold gains are not primarily driven by trivial identifier retention.

\subsubsection{Supplementary Setting}

Role reversal transforms the complete attribution: re-solving becomes negative under the weaker reviewer, strong-draft content turns positive and offsets this capability loss, and scaffold remains pair-dependent. Together with harmful primary-setting content, this shows that draft utility depends on whether the incoming artifact offers usable structure and semantics rather than brittle baggage. Across both tasks, the two heterogeneous pairs reproduce the same regimes, identifying task structure and draft quality, rather than model identity alone, as the organizing axes of collaborative revision.

\subsection{Mechanistic Case Studies of Second-Pass Behavior}
\label{sec:case_studies}

\begin{figure}[H]
\vspace{-5pt}
    \begin{center}
    \includegraphics[width=0.95\linewidth]{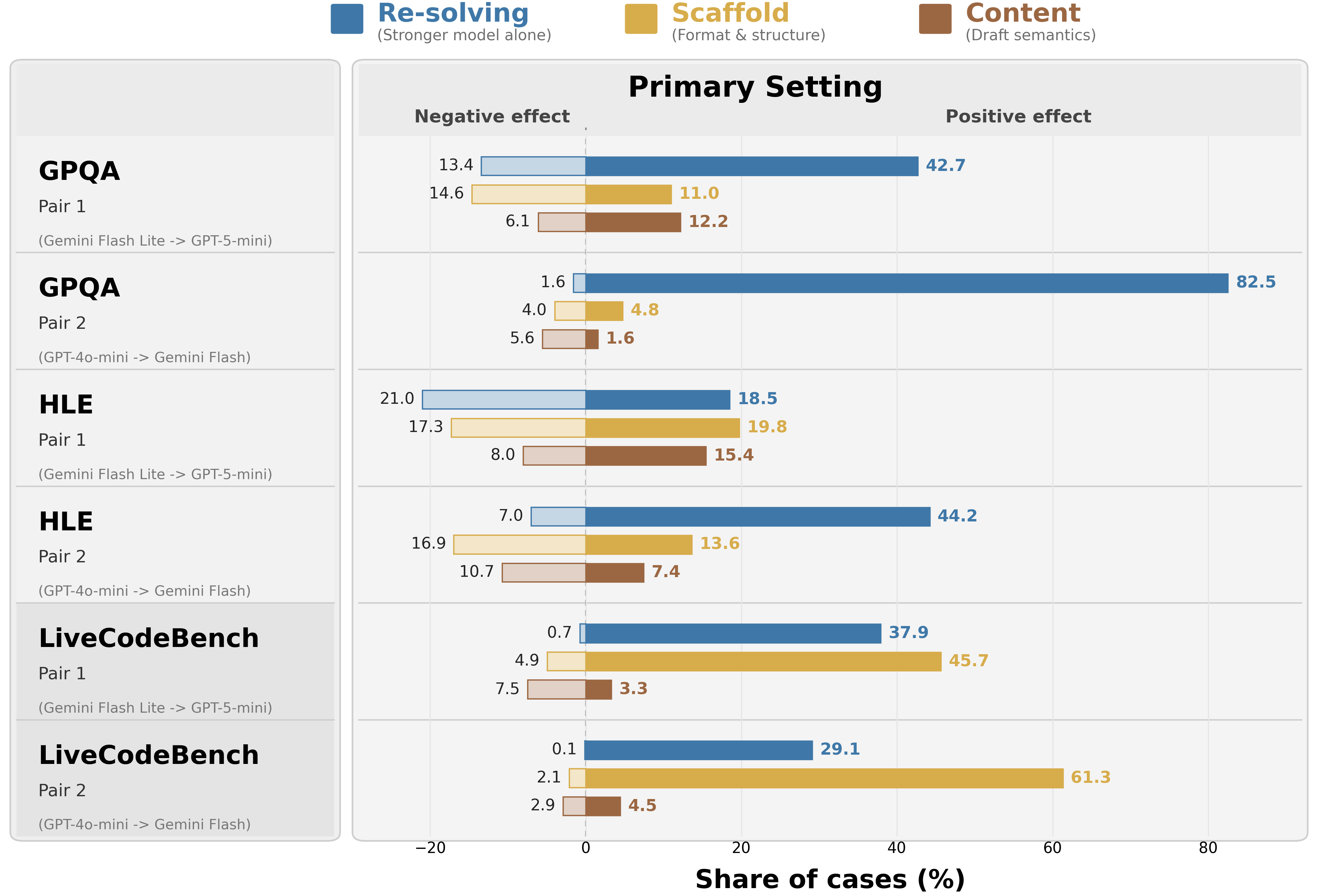}
    \end{center}
    \caption{Mechanism-level decomposition of second-pass outcomes in the primary setting. On MCQ, diagnostic cases are dominated by the re-solving family; on LiveCodeBench, they are dominated by scaffold-positive cases.}
    \label{fig:primary_taxonomy_summary}
    \vspace{-4pt}
\end{figure}
\vspace{-10pt}

The aggregate decompositions above identify \emph{which} component dominates on average; this subsection asks \emph{how} those effects appear in individual examples.

\paragraph{Per-question diagnostic taxonomy.}
Each question induces an outcome tuple $(x_1x_2x_3x_4)$, where \cm\ denotes correct and \xm\ denotes incorrect. We exclude the two non-diagnostic extremes, \mbox{(\xm\xm\xm\xm)} and \mbox{(\cm\cm\cm\cm)}, and assign the remaining 14 patterns to exactly one family using a priority rule. We first check \textbf{Content}, because $x_2$ vs.\ $x_4$ is the cleanest contrast where the prompt is identical and only the draft content differs. Among the remaining cases, we then check \textbf{Scaffold}, asking whether the null scaffold changes the outcome relative to unassisted re-solving. Residual cases are assigned to \textbf{Re-solving}, where the reviewer behaves the same regardless of the draft and only differs from the generator baseline. Full 16-way counts appear in Appendix Table~\ref{tab:primary_pair_pattern_counts}.

\begin{itemize}
    \item \textbf{Content} ($x_2 \neq x_4$): the real draft changes the outcome relative to the matched null scaffold. Positive means $x_2$ is correct while $x_4$ is incorrect, e.g., \mbox{(\xm\cm\xm\xm)}. Negative means $x_2$ is incorrect while $x_4$ is correct, e.g., \mbox{(\cm\xm\cm\cm)}.

    \item \textbf{Scaffold} ($x_2 = x_4$ but $x_3 \neq x_4$): draft content is inert, but review framing with the null scaffold changes the outcome relative to pure re-solving. Positive means $x_3$ is incorrect while $x_4$ is correct, e.g., \mbox{(\xm\cm\xm\cm)}. Negative means $x_3$ is correct while $x_4$ is incorrect, e.g., \mbox{(\xm\xm\cm\xm)}.

    \item \textbf{Re-solving} ($x_2 = x_4 = x_3$ but $x_1 \neq x_2$): neither draft content nor review framing matters; only the reviewer-generator capability gap remains. Positive means the reviewer succeeds where the generator fails, e.g., \mbox{(\xm\cm\cm\cm)}. Negative / drift means the generator is correct but the reviewer independently errs, e.g., \mbox{(\cm\xm\xm\xm)}.
\end{itemize}
 
Figure~\ref{fig:primary_taxonomy_summary} shows that these families distribute very differently across task types. On MCQ, diagnostic cases are dominated by re-solving, whereas on LiveCodeBench they shift toward scaffold-positive cases and content-negative cases. We next examine representative examples of these contrasting behaviors.

\subsubsection{MCQ: Re-solving and second-pass drift}

\begin{figure}[h]
\vspace{-5pt}
    \begin{center}
    \includegraphics[width=1.0\linewidth]{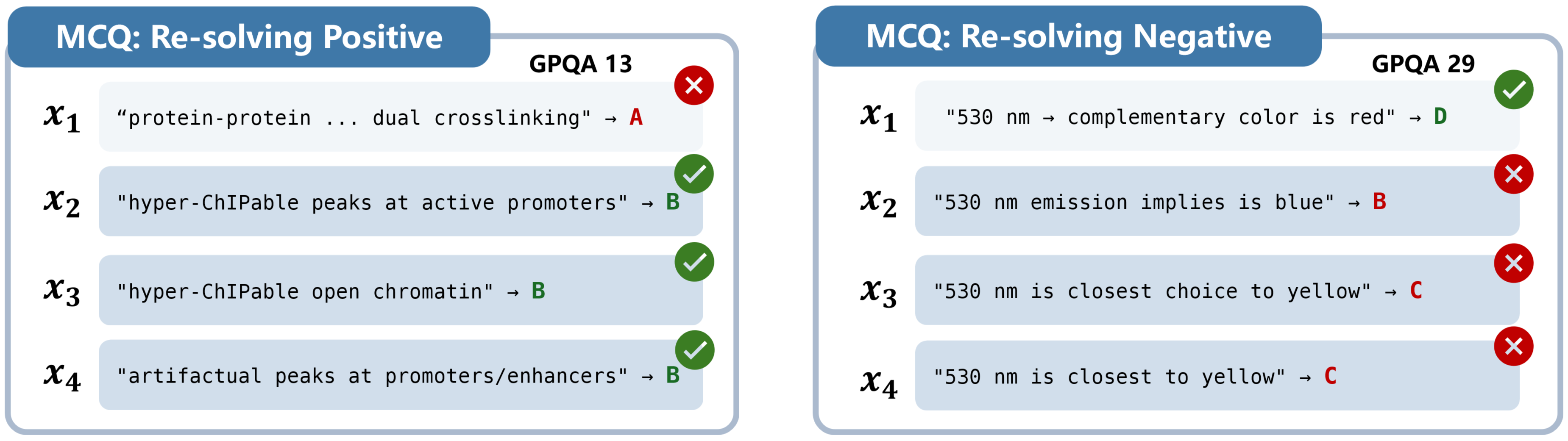}
    \end{center}
    \caption{Representative MCQ cases. Left: positive re-solving, where the reviewer abandons the weak premise and reaches the correct answer independently. Right: negative re-solving, where the second pass overrides a correct weak answer and drifts to an incorrect one.}
    \label{fig:plot3_mcq_mechanism}
    \vspace{-8pt}
\end{figure}

\paragraph{Positive Re-solving.}
In the left panel of Figure~\ref{fig:plot3_mcq_mechanism}, the weak generator ($x_1$) starts from an incorrect premise and answers incorrectly, while the reviewer-side conditions reach the correct answer through a different conceptual route. The key point is that the reviewer does not refine the weak draft. Instead, it abandons the draft and reconstructs the answer from scratch, so the gain comes from the stronger model's independent reasoning rather than from reliable semantic value in the weak draft.

\vspace{-8pt}

\paragraph{Negative Re-solving.}
The right panel shows the complementary failure mode. Here the weak generator is already correct, but all reviewer-side conditions replace that correct answer with new, incorrect trajectories. Rather than preserving a correct baseline, the second pass re-engages the question and drifts away from it. On MCQ, then, the same replacement behavior that can rescue a weak answer can also overwrite a correct one.

\subsubsection{Code: Artifact-level anchoring and scaffold repair}

On code-generation tasks, the second pass behaves less like reasoning revision and more like artifact repair. Weak drafts can anchor the reviewer to brittle implementations, while scaffolds can improve performance by inviting structured completion.

\begin{figure}[!h]
\vspace{-5pt}
    \begin{center}
    \includegraphics[width=1.0\linewidth]{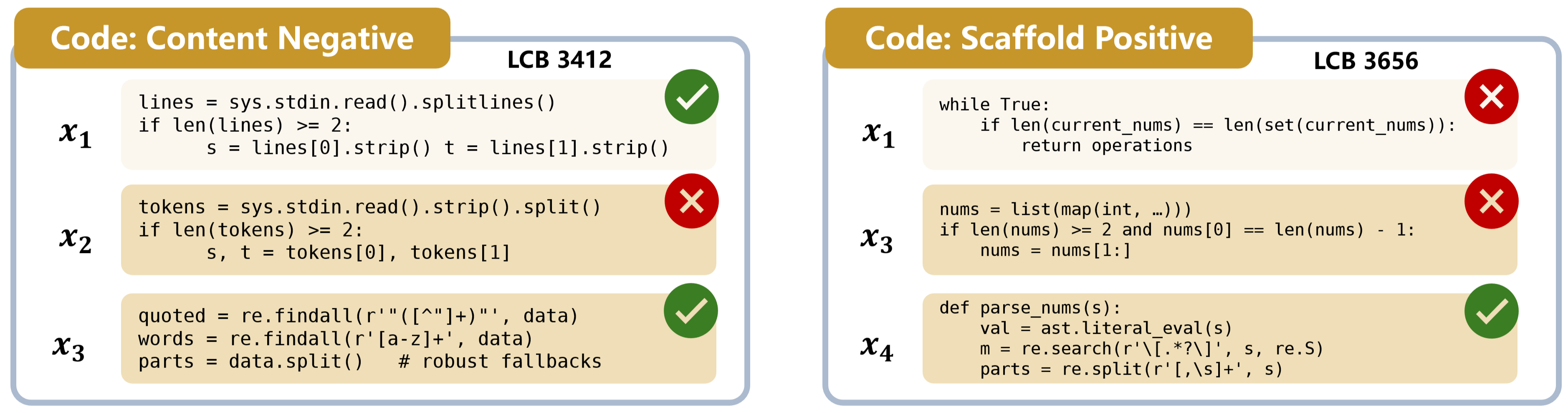}
    \end{center}
    \caption{Representative code cases. Left: negative content, where a draft anchors revision to a brittle artifact. Right: positive scaffold, where scaffold supports a robust solution.}    \label{fig:plot4_code_mechanism}
    \vspace{-8pt}
\end{figure}

\paragraph{Negative Content.}
In the left panel of Figure~\ref{fig:plot4_code_mechanism}, the harm does not come from the weak draft's high-level idea alone, but from its concrete implementation form. The weak draft exposes brittle parsing and wrapper decisions, and the reviewer remains anchored to this local structure instead of rebuilding a more robust solution. The result is a submission that inherits implementation fragility even when the stronger reviewer could, in principle, solve the problem correctly on its own.

\paragraph{Positive Scaffold.}
The right panel clarifies why the null scaffold can help. The scaffold carries no algorithmic solution, but it does provide a syntactically valid, code-shaped object that invites completion. In the review framing, this shifts the task away from free-form solving and toward submission repair including finishing the wrapper, normalizing I/O structure, and producing a complete executable artifact. This kind of structural benefit has no close analogue on MCQ.

\vspace{-8pt}

\paragraph{Why the mechanisms diverge across tasks.}
We hypothesize that this divergence reflects differences in output-space structure. MCQ has a constrained answer space, so a strong reviewer can often override a bad prefix and reconstruct the answer independently. However, code generation is far more open-ended. As later tokens depend heavily on earlier structural choices, the reviewer is more vulnerable to anchoring by a weak draft and more likely to benefit from a scaffold that frames the task as repair rather than fresh generation.

\FloatBarrier
\subsection{Broader Implications for Collaborative AI Systems}
\label{sec:broader_implications}

The framework supplies a reusable unit of analysis for collaborative intelligence: the \emph{handoff}. Whenever an upstream model or agent passes an intermediate artifact to a downstream component, the resulting gain can be attributed to downstream capability, structural scaffolding, and transmitted semantic content. The same counterfactuals can therefore be instantiated beyond pairwise revision in model cascades, planner--executor--verifier architectures, debate and reflection systems, deep-research agents, continual skill learners, code-repair workflows, and closed-loop scientific pipelines. These systems pass retrieved evidence or procedural skills across steps~\citep{coelho2026deepresearchgym,ning2026agenticsearch,zhong2026skilllearnbench}, as well as executable artifacts and measured lineage~\citep{ning2026autoresearch,ning2026closedloop}.

This view distinguishes genuine collaboration from nominal collaboration that primarily adds downstream computation. Whether applied to individual handoffs or to individual edges in larger agent graphs, the decomposition can guide dynamic routing, agent selection and pruning, communication compression, scaffold design, evidence allocation, trajectory diversification, and compute allocation. Corpus--model trade-offs and diverse query initialization show that gains can come from better evidence or exploration, not only from larger models or homogeneous parallel rollouts~\citep{ning2026lessllm,murali2026divinit}. Rather than ranking systems only by final accuracy, component-level evaluation can measure which handoffs transfer semantics, supply useful structure, or merely trigger re-solving. The framework is thus a general methodology for designing and auditing modular AI systems.

\section{Conclusion}
\label{sec:conclusion}

Our decomposition framework changes how gains in multi-LLM collaboration can be interpreted. Rather than treating end-to-end improvement as evidence of successful information transfer, it attributes the gain to re-solving, scaffold, and content. On knowledge-intensive MCQ tasks in the primary weak$\rightarrow$strong setting, apparent improvements are overwhelmingly driven by the stronger model independently re-solving the problem, with content effects remaining near zero. On code generation, by contrast, semantically empty scaffolds provide substantial value while weak-draft content can actively degrade performance.

The role-reversed experiments complete this picture: re-solving becomes costly when the reviewer is weaker, whereas high-quality drafts transfer useful capability downstream. The same qualitative regimes recur across both heterogeneous model pairs. Task structure and draft quality therefore provide a more informative basis for designing collaborative inference than the blanket assumption that an additional critic or reviewer will help through correction. This perspective supports targeted routing for constrained reasoning tasks and scaffold-centered pipelines for structured generation tasks.

More broadly, our contribution is a general framework for auditing sequential computation across models and agents. Any handoff in a cascade or agent graph can be examined through downstream capability, interface structure, and upstream semantics. This diagnostic supports principled topology, communication, routing, compute allocation, training, and evaluation across multi-agent workflows.

\clearpage
\bibliography{references}
\bibliographystyle{colm2026_conference}

\clearpage
\appendix
\clearpage

\startcontents[appendix]

\section*{Contents of Appendix}
\printcontents[appendix]{}{1}[1]{}

\section{Experimental Setup and Prompt Templates}
\label{app:exp_setup}

\paragraph{Method Details}
Our experiments instantiate the same revision design with two model pairs and two role directions. In the \emph{primary} setting, Pair~1 uses Gemini Flash Lite as the generator and GPT-5-mini as the reviewer, while Pair~2 uses GPT-4o-mini as the generator and Gemini Flash as the reviewer. In the \emph{supplementary} setting, the generator and reviewer roles are swapped within each pair. Concretely, the code defines four model assignments: primary pair~1 (Gemini Flash Lite $\rightarrow$ GPT-5-mini), supplementary pair~1 (GPT-5-mini $\rightarrow$ Gemini Flash Lite), primary pair~2 (GPT-4o-mini $\rightarrow$ Gemini Flash), and supplementary pair~2 (Gemini Flash $\rightarrow$ GPT-4o-mini). The prompt logic is held fixed across all four settings and only the model assignment changes. This makes the two series byte-for-byte identical in prompt structure and condition logic, isolating model-pair effects from implementation artifacts.

Within each setting, we evaluate four main conditions. $x_1$ is the generator-only baseline, where the generator answers the question directly. $x_2$ is standard cross-model revision, where the reviewer receives the question together with the real draft produced under $x_1$. $x_3$ is a pure re-solving control, where the reviewer receives the question only and answers directly with no review framing. $x_4$ is a scaffold control, where the reviewer receives the same review prompt as in $x_2$ but the draft is replaced with a semantically null placeholder. As implemented in the experimental script, $x_2$ and $x_4$ share the \emph{exact same} critique-prompt template; only the \texttt{draft} argument differs. Likewise, $x_3$ uses the same direct-answer template as $x_1$, ensuring that any $x_2 > x_3$ gap cannot be attributed to privileged prompt framing. No critique prompt includes source attribution indicating which model produced the draft.

For the coding task, we additionally define an ablation condition $x_5$, used only to test possible identifier leakage in the null scaffold. Compared with $x_4$, the only change is that the placeholder no longer attempts to preserve any task-specific function name or format signal. This produces a maximally neutral scaffold and allows us to distinguish the effect of format-matching from the effect of weak-draft content itself.

The semantic-null draft in $x_4$ is constructed to preserve output format while carrying no task-specific semantics. For MCQ, the placeholder always follows the same three-line structure (\texttt{Reason 1}, \texttt{Reason 2}, \texttt{Answer}) as a real model output. The answer letter is assigned deterministically by \texttt{MD5(question)[0] \% 4}, which makes the placeholder reproducible across runs and machines while distributing letters uniformly over \texttt{A--D}. For code, the null draft is a syntactically valid Python stub. Function-level problems receive a placeholder function whose name is extracted from the problem statement via regex so that the signature matches the required identifier exactly; stdin/stdout problems receive a \texttt{main()} scaffold with a \texttt{pass} body. This design avoids contaminating the null baseline with accidental content signals such as a wrong function name. By contrast, $x_5$ uses a fixed generic stub (e.g., \texttt{def solution(*args, **kwargs): ...}) with no task-specific lexical cues.

The implementation also uses a selective cache to ensure clean comparisons. Generator steps ($x_1$ and $x_1^r$) are cached and reused by downstream conditions so that all revision and scaffold variants for a given question operate on the same generator draft. Critique and re-solving steps ($x_2$, $x_3$, $x_4$, and $x_5$, together with their supplementary counterparts) always bypass the cache and are issued as fresh API calls. The cache key is the tuple \texttt{(model\_key, tag, prompt)}, which prevents collisions between different model roles or condition types even when prompts share overlapping content.

\paragraph{Prompt Templates}
We separate the prompt templates by function rather than placing all prompts in a single omnibus box. Direct-answer prompts are used for $x_1$ and $x_3$, review prompts are used for $x_2$, $x_4$, and $x_5$, and null-draft templates are listed separately because they define the scaffold and true-null controls.

\begin{figure}[H]
\centering

\begin{minipage}[t]{0.49\textwidth}
\begin{NewBoxFloat}[ResolveBlue]{MCQ direct-answer prompt ($x_1$, $x_3$)}{box:mcq_direct_prompt}
{\scriptsize\ttfamily\raggedright
Answer the following multiple-choice question.\par
Reply in this EXACT format (no other text):\par
Reason 1: <one short reason supporting your answer>\par
Reason 2: <one short reason why the most plausible alternative is wrong>\par
Answer: <letter>\par
\par
\{QUESTION\}\par
}
\end{NewBoxFloat}
\end{minipage}
\hfill
\begin{minipage}[t]{0.49\textwidth}
\begin{NewBoxFloat}[ScaffoldGold]{Code direct-answer prompt ($x_1$, $x_3$)}{box:code_direct_prompt}
{\scriptsize\ttfamily\raggedright
\textbf{Function-level tasks}\par
Write a Python function that solves the following problem.\par
Return only the function code, no explanation.\par
\par
Problem:\par
\{QUESTION\}\par

\par\smallskip
\textbf{stdin/stdout tasks}\par
Solve the following competitive programming problem in Python.\par
Your solution MUST read input from stdin and write output to stdout.\par
Return only the complete Python solution, no explanation.\par
\par
Problem:\par
\{QUESTION\}\par
}
\end{NewBoxFloat}
\end{minipage}

\end{figure}

\begin{figure}[H]
\centering

\begin{minipage}[t]{0.49\textwidth}
\begin{NewBoxFloat}[ResolveBlue]{MCQ review prompt ($x_2$, $x_4$)}{box:mcq_review_prompt}
{\scriptsize\ttfamily\raggedright
Here is an answer to the following multiple-choice question:\par
\par
Question:\par
\{QUESTION\}\par
\par
Answer attempt:\par
\{DRAFT\}\par
\par
Review the answer and reasons above. Identify any errors in the reasoning or the chosen letter, then provide your own best answer.\par
Reply in this EXACT format (no other text):\par
Reason 1: <one short reason supporting your answer>\par
Reason 2: <one short reason why the most plausible alternative is wrong>\par
Answer: <letter>\par
}
\end{NewBoxFloat}
\end{minipage}
\hfill
\begin{minipage}[t]{0.49\textwidth}
\begin{NewBoxFloat}[ScaffoldGold]{Code review prompt ($x_2$, $x_4$, $x_5$)}{box:code_review_prompt}
{\scriptsize\ttfamily\raggedright
Here is a Python solution to the following problem:\par
\par
Problem:\par
\{QUESTION\}\par
\par
Solution:\par
\{DRAFT\}\par
\par
Review this solution carefully, identify any errors or suboptimal logic, and provide a corrected and improved version.\par
For function-level tasks: Return only the improved function code, no explanation.\par
For stdin/stdout tasks: Return only the complete improved Python solution (reads from stdin, writes to stdout), no explanation.\par
}
\end{NewBoxFloat}
\end{minipage}

\end{figure}

\begin{figure}[H]
\centering

\begin{minipage}{0.78\linewidth}
\begin{NewBoxFloat}[ResolveBlue]{MCQ null draft for $x_4$}{box:mcq_null_prompt}
{\scriptsize\ttfamily\raggedright
Reason 1: This answer considers several possibilities but does not address the current question.\par
Reason 2: The reasoning pattern is generic and may not be relevant here.\par
Answer: X\par
\par
X is assigned deterministically by MD5(question)[0] \% 4.\par
}
\end{NewBoxFloat}
\end{minipage}

\vspace{0.7em}

\begin{minipage}{0.78\linewidth}
\begin{NewBoxFloat}[ScaffoldGold]{Code null scaffold for $x_4$}{box:code_null_prompt}
{\scriptsize\ttfamily\raggedright
\textbf{Function-level}\par
def \{FUNC\_NAME\}(*args, **kwargs):\par
\ \ \ \# placeholder implementation\par
\ \ \ \# structure only; not related to the current task\par
\ \ \ pass\par

\par\smallskip
\par\smallskip
\par\smallskip
\textbf{stdin/stdout}\par
import sys\par
\par
def main():\par
\ \ \ \# placeholder implementation\par
\ \ \ \# structure only; not related to the current task\par
\ \ \ pass\par
\par
main()\par
}
\end{NewBoxFloat}
\end{minipage}

\vspace{0.7em}

\begin{minipage}{0.78\linewidth}
\begin{NewBoxFloat}[ContentBrown]{Code true-null draft for $x_5$}{box:code_truenull_prompt}
{\scriptsize\ttfamily\raggedright
def solution(*args, **kwargs):\par
\ \ \ """Placeholder draft with no task-specific information."""\par
\ \ \ raise NotImplementedError\par
}
\end{NewBoxFloat}
\end{minipage}

\end{figure}

\paragraph{Hyperparameters}
The LLM wrapper defines four model-role keys and maps them to concrete API model names: \texttt{gpt\_strong} $\rightarrow$ \texttt{gpt-5-mini}, \texttt{gpt\_weak} $\rightarrow$ \texttt{gpt-4o-mini}, \texttt{gemini\_strong} $\rightarrow$ \texttt{gemini-3-flash-preview}, and \texttt{gemini\_weak} $\rightarrow$ \texttt{gemini-3.1-flash-lite-preview}. OpenAI calls are issued through \texttt{chat.completions.create} with a single user message, and Gemini calls are issued through \texttt{generate\_content} with the prompt as the \texttt{contents} field.

Importantly, the wrapper does \emph{not} explicitly set decoding controls such as temperature, top-\emph{p}, max tokens, or random seed. As a result, all runs use the providers' default generation settings. The only explicit execution-time control exposed in the wrapper is \texttt{use\_cache}. When \texttt{use\_cache=True}, the system reads from a persistent cache if a matching entry already exists and otherwise writes the result on a cache miss. When \texttt{use\_cache=False}, the system always makes a fresh API call and does not read from or write to the cache. In our experiments, generator steps always use caching so that all downstream conditions share the exact same draft for a given question, whereas critique and re-solving steps always bypass the cache to avoid reusing reviewer outputs. The persistent cache file is stored as \texttt{cache.pkl}, and the wrapper reads API credentials from the environment variables \texttt{OPENAI\_API\_KEY} and \texttt{GEMINI\_API\_KEY}. The cache is guarded by a thread lock, so reads and writes are synchronized even if API calls are issued concurrently.

\section{Complete Experimental Results}
\label{app:full_results}

\subsection{Full Accuracy Table}

Table~\ref{tab:full_accuracy} reports accuracy for all 48 experimental conditions across both experiment pairs.

\begin{table}[H]
\centering
\renewcommand{\arraystretch}{1.3}
\caption{Accuracy for all conditions. Format: correct count (percentage). Pair 1 uses Gemini Flash Lite as generator and GPT-5-mini as reviewer. Pair 2 uses GPT-4o-mini as generator and Gemini Flash as reviewer.}
\label{tab:full_accuracy}
\small
\begin{tabular}{lllcccc}
\toprule
Dataset & Setting & Pair & $x_1$ & $x_2$ & $x_3$ & $x_4$ \\
\midrule
\multirow{4}{*}{GPQA (198)}
  & \multirow{2}{*}{Primary}
    & Pair 1 & 122 (61.6\%) & 157 (79.3\%) & 151 (76.3\%) & 148 (74.7\%) \\
  &   & Pair 2 & 66 (33.3\%)  & 173 (87.4\%) & 178 (89.9\%) & 179 (90.4\%) \\
\cmidrule(lr){2-7}
  & \multirow{2}{*}{Supplem.}
    & Pair 1 & 153 (77.3\%) & 143 (72.2\%) & 124 (62.6\%) & 114 (57.6\%) \\
  &   & Pair 2 & 178 (89.9\%) & 114 (57.6\%) & 68 (34.3\%)  & 61 (30.8\%) \\
\midrule
\multirow{4}{*}{HLE (451)}
  & \multirow{2}{*}{Primary}
    & Pair 1 & 57 (12.6\%)  & 91 (20.2\%)  & 79 (17.5\%)  & 83 (18.4\%) \\
  &   & Pair 2 & 51 (11.3\%)  & 169 (37.5\%) & 188 (41.7\%) & 180 (39.9\%) \\
\cmidrule(lr){2-7}
  & \multirow{2}{*}{Supplem.}
    & Pair 1 & 81 (18.0\%)  & 74 (16.4\%)  & 64 (14.2\%)  & 61 (13.5\%) \\
  &   & Pair 2 & 188 (41.7\%) & 99 (22.0\%)  & 50 (11.1\%)  & 45 (10.0\%) \\
\midrule
\multirow{4}{*}{LCB (1054)}
  & \multirow{2}{*}{Primary}
    & Pair 1 & 533 (50.6\%) & 884 (83.9\%) & 644 (61.1\%) & 917 (87.0\%) \\
  &   & Pair 2 & 300 (28.5\%) & 823 (78.1\%) & 454 (43.1\%) & 906 (86.0\%) \\
\cmidrule(lr){2-7}
  & \multirow{2}{*}{Supplem.}
    & Pair 1 & 627 (59.5\%) & 681 (64.6\%) & 550 (52.2\%) & 647 (61.4\%) \\
  &   & Pair 2 & 461 (43.7\%) & 544 (51.6\%) & 297 (28.2\%) & 279 (26.5\%) \\
\bottomrule
\end{tabular}
\end{table}

\subsection{Complete Effect Decomposition}

Table~\ref{tab:full_decomp} reports all effect sizes in percentage points.

\begin{table}[H]
\centering
\caption{Effect decomposition for all datasets, settings, and model pairs. All values are in percentage points. Re-solving $= x_3-x_1$; Scaffold $= x_4-x_3$; Content $= x_2-x_4$; Total $= x_2-x_1$.}
\label{tab:full_decomp}
\small
\begin{tabular}{cccrrrr}
\toprule
Dataset & Setting & Pair & Total & Re-solv. & Scaffold & Content \\
\midrule
\multirow{4}{*}{GPQA}
  & \multirow{2}{*}{Primary}
    & Pair 1 & $+17.7$ & $+14.6$ & $-1.5$ & $+4.5$ \\
  &   & Pair 2 & $+54.0$ & $+56.6$ & $+0.5$ & $-3.0$ \\
\cmidrule(lr){2-7}
  & \multirow{2}{*}{Supplem.}
    & Pair 1 & $-5.1$  & $-14.6$ & $-5.1$ & $+14.6$ \\
  &   & Pair 2 & $-32.3$ & $-55.6$ & $-3.5$ & $+26.8$ \\
\midrule
\multirow{4}{*}{HLE}
  & \multirow{2}{*}{Primary}
    & Pair 1 & $+7.5$  & $+4.9$  & $+0.9$ & $+1.8$ \\
  &   & Pair 2 & $+26.2$ & $+30.4$ & $-1.8$ & $-2.4$ \\
\cmidrule(lr){2-7}
  & \multirow{2}{*}{Supplem.}
    & Pair 1 & $-1.6$  & $-3.8$  & $-0.7$ & $+2.9$ \\
  &   & Pair 2 & $-19.7$ & $-30.6$ & $-1.1$ & $+12.0$ \\
\midrule
\multirow{4}{*}{LCB}
  & \multirow{2}{*}{Primary}
    & Pair 1 & $+33.3$ & $+10.5$ & $+25.9$ & $-3.1$ \\
  &   & Pair 2 & $+49.6$ & $+14.6$ & $+42.9$ & $-7.9$ \\
\cmidrule(lr){2-7}
  & \multirow{2}{*}{Supplem.}
    & Pair 1 & $+5.1$  & $-7.3$  & $+9.2$  & $+3.2$ \\
  &   & Pair 2 & $+7.9$  & $-15.6$ & $-1.7$  & $+25.1$ \\
\bottomrule
\end{tabular}
\end{table}

\subsection{McNemar's Test Results}

Tables~\ref{tab:mcnemar_gpqa}--\ref{tab:mcnemar_lcb} report full McNemar's test statistics for all key comparisons.
Significance: $^{*}p<.05$, $^{**}p<.01$, $^{***}p<.001$.
$n_{\cm\xm}$ : condition A correct, B wrong. $n_{\xm\cm}$: condition A wrong, B correct.

\begin{table}[H]
\centering
\renewcommand{\arraystretch}{1.3}
\caption{McNemar's tests on GPQA (two-tailed, Yates correction).}
\label{tab:mcnemar_gpqa}
\small
\resizebox{\linewidth}{!}{
\begin{tabular}{ccc ccccccc}
\toprule
Setting & Pair & Comparison & Acc\,A & Acc\,B & $\Delta$ & $n_{\cm\xm}$ & $n_{\xm\cm}$ & $\chi^2$ & $p$ \\
\midrule
\multirow{6}{*}{Primary}
  & \multirow{3}{*}{Pair 1}
    & Content ($x_2 - x_4$)         & 79.3 & 74.7 & $+4.5$  & 16 & 7   & 2.78  & .095\,ns \\
  &   & Scaffold ($x_4 - x_3$)      & 74.7 & 76.3 & $-1.5$  & 9  & 12  & 0.19  & .663\,ns \\
  &   & Re-solving ($x_3 - x_1$)    & 76.3 & 61.6 & $+14.6$ & 45 & 16  & 12.85 & .0003*** \\
\cmidrule(lr){2-10}
  & \multirow{3}{*}{Pair 2}
    & Content ($x_2 - x_4$)         & 87.4 & 90.4 & $-3.0$  & 4  & 10  & 1.79  & .181\,ns \\
  &   & Scaffold ($x_4 - x_3$)      & 90.4 & 89.9 & $+0.5$  & 6  & 5   & 0.00  & 1.00\,ns \\
  &   & Re-solving ($x_3 - x_1$)    & 89.9 & 33.3 & $+56.6$ & 114 & 2  & 106.2 & $<$.001*** \\
\midrule
\multirow{6}{*}{Supplem.}
  & \multirow{3}{*}{Pair 1}
    & Content ($x_{2r} - x_{4r}$)   & 72.2 & 57.6 & $+14.6$ & 42 & 13 & 14.26 & .0002*** \\
  &   & Scaffold ($x_{4r} - x_{3r}$)& 57.6 & 62.6 & $-5.1$  & 22 & 32 & 1.50  & .221\,ns \\
  &   & Re-solving ($x_{3r} - x_{1r}$) & 62.6 & 77.3 & $-14.6$ & 16 & 45 & 12.85 & .0003*** \\
\cmidrule(lr){2-10}
  & \multirow{3}{*}{Pair 2}
    & Content ($x_{2r} - x_{4r}$)   & 57.6 & 30.8 & $+26.8$ & 64  & 11 & 36.05 & $<$.001*** \\
  &   & Scaffold ($x_{4r} - x_{3r}$)& 30.8 & 34.3 & $-3.5$  & 24  & 31 & 0.65  & .418\,ns \\
  &   & Re-solving ($x_{3r} - x_{1r}$) & 34.3 & 89.9 & $-55.6$ & 2  & 112 & 104.2 & $<$.001*** \\
\bottomrule
\end{tabular}
}
\end{table}

\begin{table}[H]
\centering
\renewcommand{\arraystretch}{1.3}
\caption{McNemar's tests on HLE (two-tailed, Yates correction).}
\label{tab:mcnemar_hle}
\small
\resizebox{\linewidth}{!}{
\begin{tabular}{ccc ccccccc}
\toprule
Setting & Pair & Comparison & Acc\,A & Acc\,B & $\Delta$ & $n_{\cm\xm}$ & $n_{\xm\cm}$ & $\chi^2$ & $p$ \\
\midrule
\multirow{6}{*}{Primary}
  & \multirow{3}{*}{Pair 1}
    & Content ($x_2 - x_4$)         & 20.2 & 18.4 & $+1.8$  & 44 & 36  & 0.61  & .434\,ns \\
  &   & Scaffold ($x_4 - x_3$)      & 18.4 & 17.5 & $+0.9$  & 32 & 28  & 0.15  & .699\,ns \\
  &   & Re-solving ($x_3 - x_1$)    & 17.5 & 12.6 & $+4.9$  & 64 & 42  & 4.16  & .041*   \\
\cmidrule(lr){2-10}
  & \multirow{3}{*}{Pair 2}
    & Content ($x_2 - x_4$)         & 37.5 & 39.9 & $-2.4$  & 33 & 44  & 1.30  & .254\,ns \\
  &   & Scaffold ($x_4 - x_3$)      & 39.9 & 41.7 & $-1.8$  & 33 & 41  & 0.66  & .416\,ns \\
  &   & Re-solving ($x_3 - x_1$)    & 41.7 & 11.3 & $+30.4$ & 166 & 29 & 94.85 & $<$.001*** \\
\midrule
\multirow{6}{*}{Supplem.}
  & \multirow{3}{*}{Pair 1}
    & Content ($x_{2r} - x_{4r}$)   & 16.4 & 13.5 & $+2.9$  & 47 & 34 & 1.78  & .182\,ns \\
  &   & Scaffold ($x_{4r} - x_{3r}$)& 13.5 & 14.2 & $-0.7$  & 32 & 35 & 0.06  & .807\,ns \\
  &   & Re-solving ($x_{3r} - x_{1r}$) & 14.2 & 18.0 & $-3.8$  & 47 & 64 & 2.31  & .129\,ns \\
\cmidrule(lr){2-10}
  & \multirow{3}{*}{Pair 2}
    & Content ($x_{2r} - x_{4r}$)   & 22.0 & 10.0 & $+12.0$ & 78  & 24 & 27.54 & $<$.001*** \\
  &   & Scaffold ($x_{4r} - x_{3r}$)& 10.0 & 11.1 & $-1.1$  & 31  & 36 & 0.24  & .625\,ns \\
  &   & Re-solving ($x_{3r} - x_{1r}$) & 11.1 & 41.7 & $-30.6$ & 28 & 166 & 96.75 & $<$.001*** \\
\bottomrule
\end{tabular}
}
\end{table}

\begin{table}[H]
\centering
\renewcommand{\arraystretch}{1.3}
\caption{McNemar's tests on LiveCodeBench (two-tailed, Yates correction).}
\label{tab:mcnemar_lcb}
\small
\resizebox{\linewidth}{!}{
\begin{tabular}{ccc ccccccc}
\toprule
Setting & Pair & Comparison & Acc\,A & Acc\,B & $\Delta$ & $n_{\cm\xm}$ & $n_{\xm\cm}$ & $\chi^2$ & $p$ \\
\midrule
\multirow{6}{*}{Primary}
  & \multirow{3}{*}{Pair 1}
    & Content ($x_2 - x_4$)         & 83.9 & 87.0 & $-3.1$  & 43  & 76  & 8.61  & .003**  \\
  &   & Scaffold ($x_4 - x_3$)      & 87.0 & 61.1 & $+25.9$ & 306 & 33 & 218.2 & $<$.001*** \\
  &   & Re-solving ($x_3 - x_1$)    & 61.1 & 50.6 & $+10.5$ & 325 & 214 & 22.45 & $<$.001*** \\
\cmidrule(lr){2-10}
  & \multirow{3}{*}{Pair 2}
    & Content ($x_2 - x_4$)         & 78.1 & 86.0 & $-7.9$  & 44 & 127  & 39.32 & $<$.001*** \\
  &   & Scaffold ($x_4 - x_3$)      & 86.0 & 43.1 & $+42.9$ & 468  & 16 & 420.3 & $<$.001*** \\
  &   & Re-solving ($x_3 - x_1$)    & 43.1 & 28.5 & $+14.6$ & 258 & 104 & 64.67 & $<$.001*** \\
\midrule
\multirow{6}{*}{Supplem.}
  & \multirow{3}{*}{Pair 1}
    & Content ($x_{2r} - x_{4r}$)   & 64.6 & 61.4 & $+3.2$  & 129  & 95 & 4.86  & .027*   \\
  &   & Scaffold ($x_{4r} - x_{3r}$)& 61.4 & 52.2 & $+9.2$  & 159  & 62 & 41.70 & $<$.001*** \\
  &   & Re-solving ($x_{3r} - x_{1r}$) & 52.2 & 59.5 & $-7.3$  & 229 & 306 & 10.80 & .001**  \\
\cmidrule(lr){2-10}
  & \multirow{3}{*}{Pair 2}
    & Content ($x_{2r} - x_{4r}$)   & 51.6 & 26.5 & $+25.1$ & 293  & 28 & 217.1 & $<$.001*** \\
  &   & Scaffold ($x_{4r} - x_{3r}$)& 26.5 & 28.2 & $-1.7$  & 39  & 57  & 3.01  & .083\,ns \\
  &   & Re-solving ($x_{3r} - x_{1r}$) & 28.2 & 43.7 & $-15.6$ & 116 & 280 & 67.09 & $<$.001*** \\
\bottomrule
\end{tabular}
}
\end{table}

\subsection{LiveCodeBench Difficulty Breakdown (Pair 1, primary)}

Table~\ref{tab:lcb_diff} shows a breakdown of the accuracy of the four conditions and content effect from different difficulty levels on LiveCodeBench.

\begin{table}[H]
\centering
\renewcommand{\arraystretch}{1.3}
\caption{Accuracy by difficulty tier on LiveCodeBench, Pair 1 primary setting. Content effect = $x_2 - x_4$.}
\label{tab:lcb_diff}
\small
\begin{tabular}{lcrcccc}
\toprule
Tier & $n$ & $x_1$ & $x_2$ & $x_3$ & $x_4$ & Content \\
\midrule
Easy   & 322 & 69.3\% & 94.1\% & 74.8\% & 94.7\% & $-$0.6\,pp \\
Medium & 382 & 47.1\% & 86.1\% & 66.8\% & 89.5\% & $-$3.4\,pp \\
Hard   & 350 & 37.1\% & 72.0\% & 42.3\% & 77.1\% & $-$5.1\,pp \\
\bottomrule
\end{tabular}
\end{table}

\subsection{Scaffold Ablation on Identifier Retention}
\label{sec:scaffold_ablation_appendix}

To test whether the gains of the scaffold control are partly driven by task-specific identifier cues, we compare the \textbf{Null Scaffold} against a more neutral \textbf{True-Null Scaffold}. The two conditions use the same review prompt and differ only in the placeholder shown to the reviewer. The Null Scaffold preserves a task-shaped placeholder, including the extracted function name when available, whereas the True-Null Scaffold replaces this with a fixed generic stub.

\begin{table}[H]
    \centering
    \begin{minipage}[c]{0.4\linewidth}
        \centering
        \includegraphics[width=0.95\linewidth]{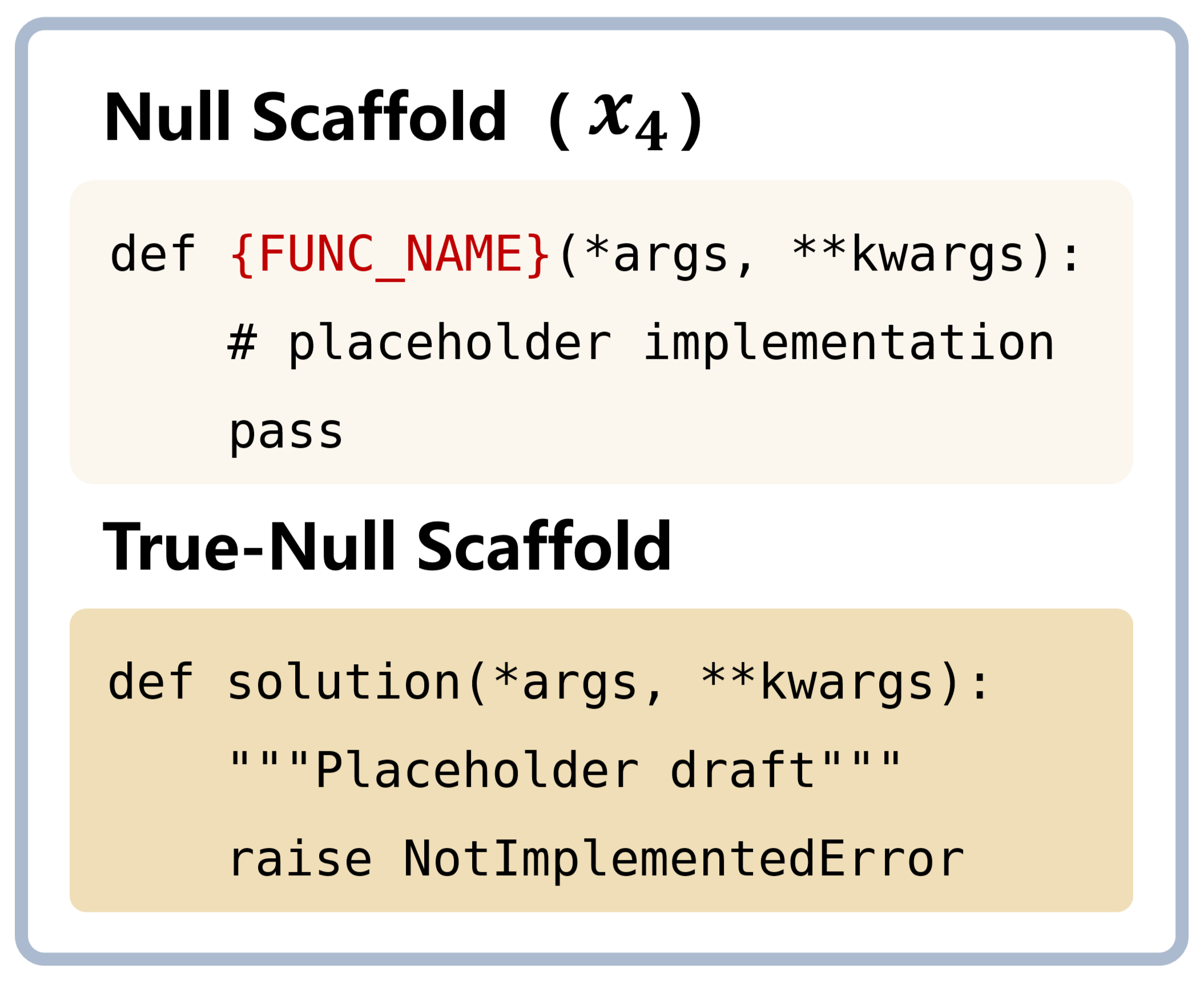}
    \end{minipage}\hfill
    \begin{minipage}[c]{0.59\linewidth}
        \centering
        \normalsize
        \renewcommand{\arraystretch}{2.0}
        \setlength{\tabcolsep}{5.5pt}
        \begin{tabular}{lccc}
        \toprule
        Pair & Null Scaffold & True-Null Scaffold & $p$ \\
        \midrule
        Pair 1 & 87.0 & 86.4 & 0.617 \\
        Pair 2 & 86.0 & 85.1 & 0.481 \\
        \bottomrule
        \end{tabular}
    \end{minipage}
    \caption{Scaffold ablation on LiveCodeBench. Left: comparison between the \textbf{Null Scaffold} and the \textbf{True-Null Scaffold}. Right: forward-setting accuracies and significance tests for their difference.}
    \label{tab:scaffold_ablation}
\end{table}

Quantitatively, the difference is negligible in both model pairs. For Pair~1, accuracy changes only from 87.0\% under the Null Scaffold to 86.4\% under the True-Null Scaffold ($p=0.617$); for Pair~2, it changes from 86.0\% to 85.1\% ($p=0.481$). We therefore do not find evidence that the strong performance of the scaffold control is primarily driven by the retained function name or by simple task-specific lexical cues.

At the same time, this ablation should be interpreted narrowly. It does not imply that all structural signals are irrelevant, nor that the two scaffold variants are behaviorally identical in every respect. Rather, it shows that the gains of the scaffold control are not reducible to trivial identifier retention alone. The main benefit appears to come from presenting the reviewer with a syntactically valid, code-shaped intermediate object under review framing, rather than from the specific lexical content of the extracted function name.

\subsection{Pattern-Family Counts}
\label{app:case_studies}
\begin{table}[H]
\centering
\renewcommand{\arraystretch}{1.2}
\caption{Pattern counts across all $2^4$ per-question outcome patterns for the two primary model pairs. Patterns are written in literal $x_1x_2x_3x_4$ order, where \cm\ denotes correct and \xm\ denotes incorrect.}
\label{tab:primary_pair_pattern_counts}
\tiny
\setlength{\tabcolsep}{3.0pt}
\resizebox{\linewidth}{!}{
\begin{tabular}{llccccccccccccccccc}
\toprule
Dataset & Pair & $n$
& \mbox{\xm\xm\xm\xm}
& \mbox{\xm\xm\xm\cm}
& \mbox{\xm\xm\cm\xm}
& \mbox{\xm\xm\cm\cm}
& \mbox{\xm\cm\xm\xm}
& \mbox{\xm\cm\xm\cm}
& \mbox{\xm\cm\cm\xm}
& \mbox{\xm\cm\cm\cm}
& \mbox{\cm\xm\xm\xm}
& \mbox{\cm\xm\xm\cm}
& \mbox{\cm\xm\cm\xm}
& \mbox{\cm\xm\cm\cm}
& \mbox{\cm\cm\xm\xm}
& \mbox{\cm\cm\xm\cm}
& \mbox{\cm\cm\cm\xm}
& \mbox{\cm\cm\cm\cm} \\
\midrule

\multirow{2}{*}{GPQA}
& P1 & 198 & 17 & 1 & 3 & 4 & 8 & 5 & 3 & 35 & 11 & 1 & 3 & 1 & 2 & 2 & 3 & 99 \\
& P2 & 198 & 10 & 3 & 3 & 6 & 2 & 3 & 1 & 104 & 2 & 0 & 0 & 1 & 0 & 0 & 1 & 62 \\
\midrule

\multirow{2}{*}{HLE}
& P1 & 451 & 281 & 22 & 8 & 12 & 19 & 8 & 14 & 30 & 34 & 1 & 1 & 1 & 6 & 1 & 5 & 8 \\
& P2 & 451 & 195 & 18 & 24 & 23 & 13 & 8 & 12 & 107 & 17 & 0 & 2 & 3 & 5 & 7 & 3 & 14 \\
\midrule

\multirow{2}{*}{LCB}
& P1 & 1054 & 77 & 21 & 10 & 45 & 15 & 83 & 16 & 254 & 5 & 5 & 2 & 5 & 7 & 197 & 5 & 307 \\
& P2 & 1054 & 97 & 104 & 5 & 21 & 34 & 261 & 10 & 222 & 1 & 1 & 1 & 1 & 0 & 102 & 0 & 194 \\
\bottomrule
\end{tabular}
}
\end{table}
Table~\ref{tab:primary_pair_pattern_counts} shows the pattern counts of all the 16 patterns for GPQA, HLE, and LiveCodeBench.

\section{Disclosure of LLM Usage}

All research contributions in this paper, including the experimental design, the four-condition decomposition framework, the choice of controls, the statistical analysis methodology, the interpretation of results, and the iterative refinement of the study, were conceived and carried out entirely by the authors.

We used LLMs in the following assistive capacities during the preparation of this work:

\begin{itemize}
    \item \textbf{Code implementation and debugging.} LLMs assisted with implementing portions of the experimental pipeline (e.g., API wrapper logic, evaluation scripts, and result aggregation) and with debugging runtime errors. All code was reviewed and validated by the authors.
    \item \textbf{Visual assets.} Certain illustrative icons in Figure~\ref{fig:decomposition_overview} were generated with the aid of an LLM-based image tool. All data-driven figures (tables, charts, and plots) were produced by author-written code from raw experimental outputs.
    \item \textbf{Writing assistance.} LLMs were used to polish prose, improve clarity, and suggest structural edits to the manuscript text. All substantive content, claims, and arguments are the authors' own.
\end{itemize}

\end{document}